\documentclass[12pt]{article}
\usepackage{graphicx}

\newcommand\pubnumber{CMS CR-2014/384}
\newcommand\pubdate{\today}

\def\napoli{Eidgen\"ossische Technische Hochschule Z\"urich} 

\def\Title#1{\begin{center} {\Large #1 } \end{center}}
\def\Author#1{\begin{center}{ \sc #1} \end{center}}
\def\Address#1{\begin{center}{ \it #1} \end{center}}

\newcommand\pubblock{\rightline{\begin{tabular}{l} \pubnumber\\
         \pubdate  \end{tabular}}}
\newenvironment{Abstract}{\begin{quotation}  }{\end{quotation}}
\newenvironment{Presented}{\begin{quotation} \begin{center} 
             PRESENTED AT\end{center}\bigskip 
      \begin{center}\begin{large}}{\end{large}\end{center} \end{quotation}}

\def\beq{\begin{equation}}
\def\eeq#1{\label{#1}\end{equation}}
\def\eeqn{\end{equation}}

\def\beqa{\begin{eqnarray}}
\def\eeqa#1{\label{#1}\end{eqnarray}}
\def\eeqan{\end{eqnarray}}

\let\bar=\overbar

\def\Dslash{\not{\hbox{\kern-4pt $D$}}}
\def\dslash{\not{\hbox{\kern-2pt $\del$}}}

\def\msb{{\bar{\ssstyle M \kern -1pt S}}}

\begin{document}
\begin{titlepage}
\pubblock

\vfill
\Title{Searches for the associated $t\bar{t}H$ production at CMS}
\vfill
\Author{ Liis Rebane for the CMS Collaboration} 
\Address{\napoli}
\vfill
\begin{Abstract}
After the recent discovery of the Higgs boson, the next important goal is to measure its properties.
Probing the Yukawa coupling of the Higgs boson to top quarks is a particularly important test of 
physics beyond the standard model. This coupling can be experimentally accessed by measuring the cross 
section of the Higgs boson production in association with a top quark pair ($t\bar{t}H$). The talk gives an 
overview of CMS results on $t\bar{t}H$ searches, using the full dataset of pp-collision data collected at
the centre of mass energies of 7 and 8 TeV. All relevant Higgs boson decay modes have been studied, including
Higgs decays to b-quarks, photons, $\tau$-leptons and multi-lepton final states. Additionally, the first Matrix Element 
Method based analysis has been carried out, that aims to further increase the sensitivity to the signal.
\end{Abstract}
\vfill
\begin{Presented}
8th International Workshop on the CKM Unitarity Triangle (CKM 2014)\\
Vienna, Austira,  September 8--12, 2014
\end{Presented}
\vfill
\end{titlepage}
\def\thefootnote{\fnsymbol{footnote}}
\setcounter{footnote}{0}

\section{Introduction}
In 2012 both CMS and ATLAS collaborations independently reported the observation of a Higgs-like particle \cite{Aad:2012tfa,Chatrchyan:2012ufa}. 
The next step for the experiments is to measure its properties and determine whether its couplings to other fundamental
particles are compatible with those expected for the standard model Higgs boson. The measurement of its Yukawa coupling to top quarks ($y_t$) is of particular interest, as its standard model value is very close to one, which could be a hint of a special role of the top quark in the electroweak symmetry breaking mechanism. Although measuring loop-induced processes like $gg\rightarrow H$ or $H\rightarrow\gamma\gamma$ provides an indirect probe of $y_t$, a direct measurement is only possible in processes, where the Higgs boson is radiated from a top quark.
The production of the Higgs boson in association with a top-quark pair ($t\bar{t}H$) has the largest cross-section of this type of reactions with a next-to-leading order prediction of about 130 fb at $\sqrt{s} = 8$ TeV. 


The searches for $t\bar{t}H$ events at CMS experiment cover all relevant decay modes of the Higgs boson, including $H\rightarrow\gamma\gamma$, $H\rightarrow b\bar{b}$, $H\rightarrow\tau_{\mathrm{had}}\tau_{\mathrm{had}}$ and multi-lepton final states \cite{Khachatryan:2014qaa}. Of the above processes $t\bar{t}H(b\bar{b})$ channel has by far the largest cross section, but it suffers from a large irreducible multi-jet background and ambiguity in the choice of b-quarks to be associated to the Higgs boson decay, leading  to a combinatorial self-background. To further increase the sensitivity of this difficult channel, a complementary analysis has been carried out, featuring the first Matrix Element Method based search for $t\bar{t}H$ events \cite{CMS:2014jga}.

\section{$t\bar{t}H$ in $H\rightarrow\gamma\gamma$ channel}
\begin{figure}[htbp]
\centering{
  \includegraphics[width=0.45\linewidth]{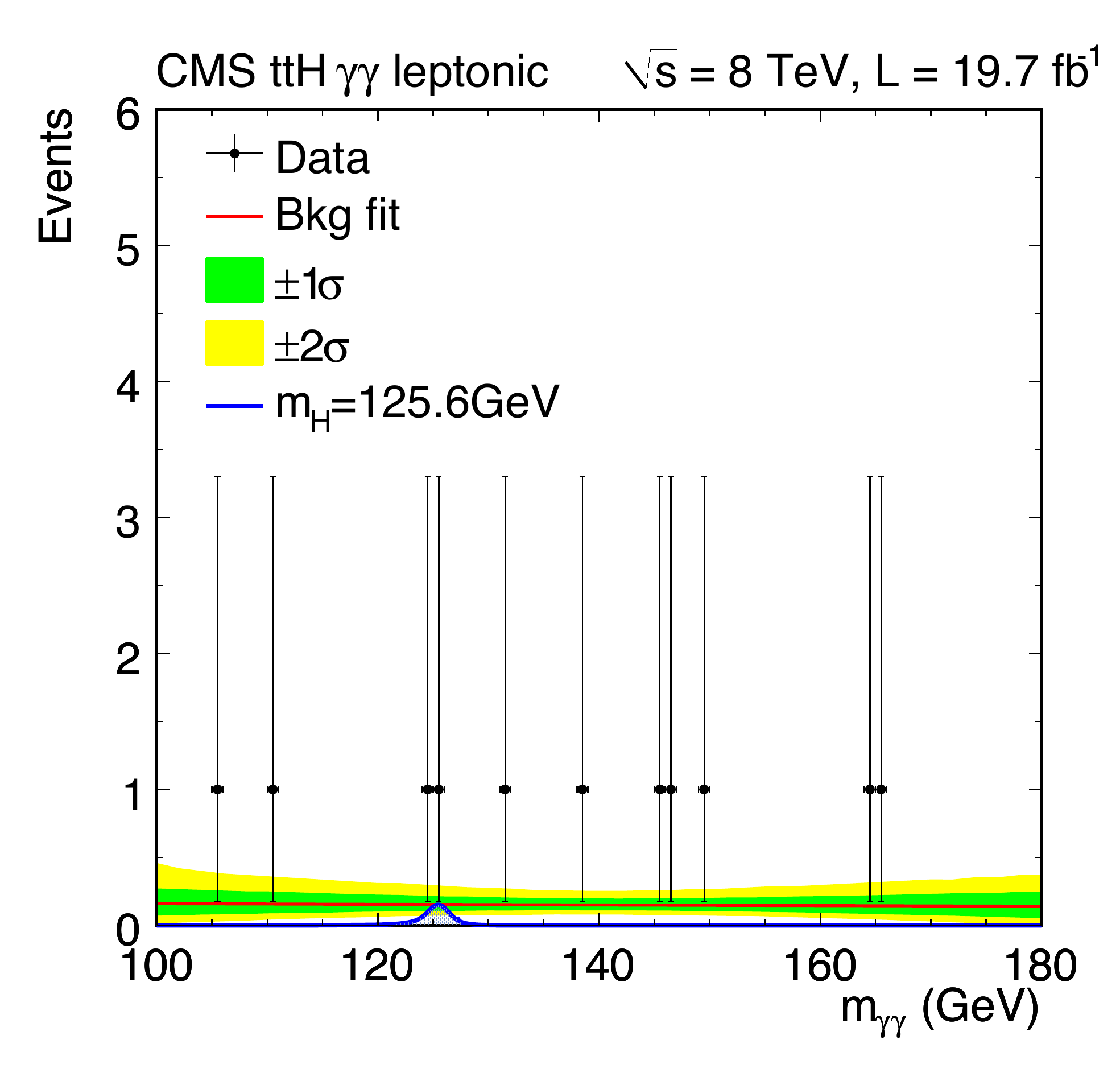}
  \includegraphics[width=0.45\linewidth]{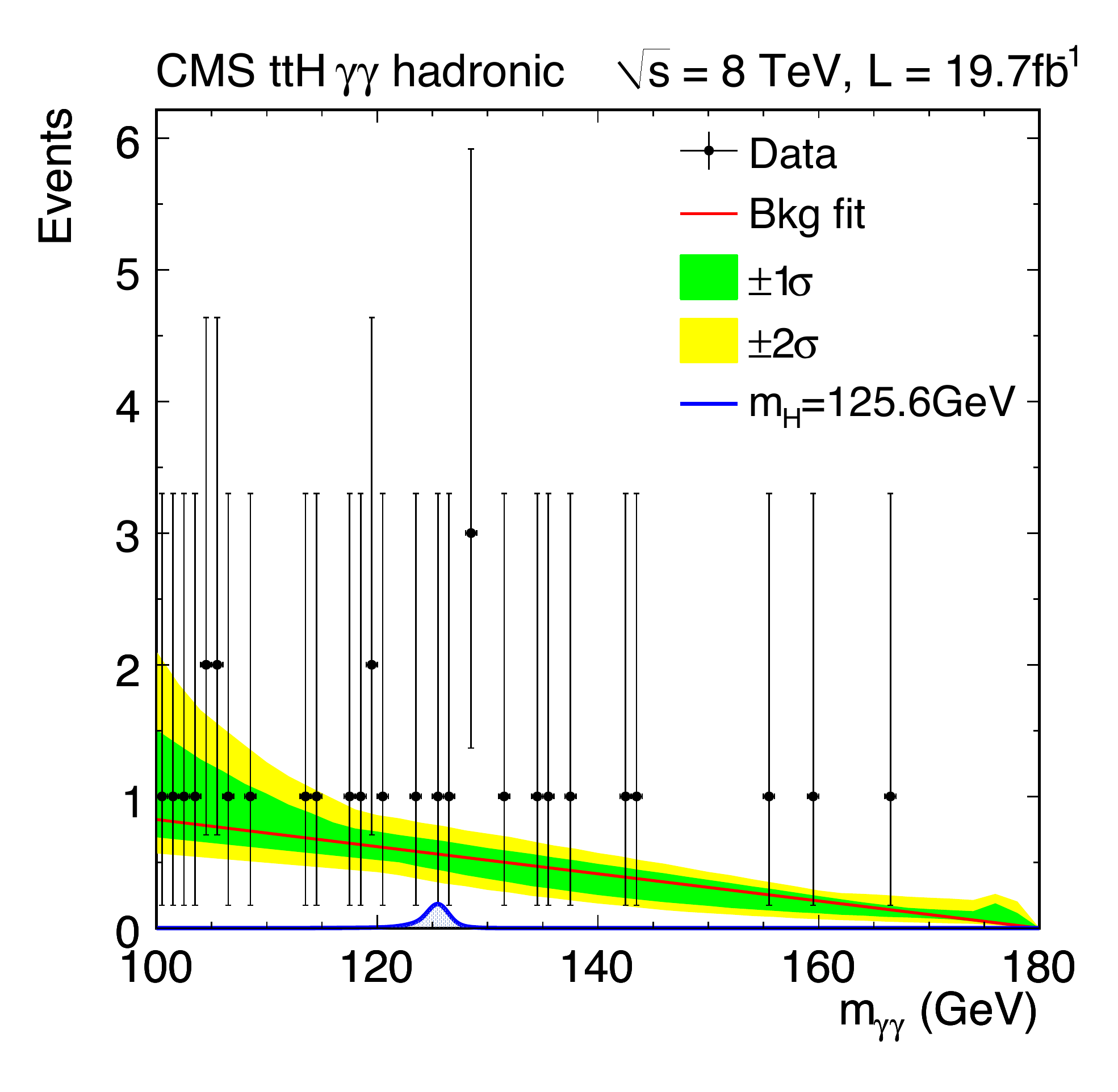}
\caption{Diphoton invariant mass distribution for selected $t\bar{t}H$ candidate events in leptonic (left) and hadronic (right) channels.}
\label{fig:photons}
}
\end{figure}

The branching fraction of $H\rightarrow\gamma\gamma$ is small and the sensitivity of this channel is strongly limited by the statistical uncertainties. However, two energetic photons from the Higgs boson decay provide a very distinct signature, allowing the $\gamma\gamma$ invariant mass to be used directly to separate a narrow signal peak from a falling distribution of the background contributions. 
As the signal events can be triggered by photons, all $t\bar{t}$ decay topologies can be considered in the analysis, including the fully hadronic decay mode. The di-photon invariant mass distribution for leptonic (left) and hadronic (right) selections is shown in Figure~\ref{fig:photons} and demonstrates good agreement between data and background estimates. 

\section{$t\bar{t}H$ in leptonic final states}
\begin{figure}[htbp]
\centering{
  \includegraphics[width=0.32\linewidth]{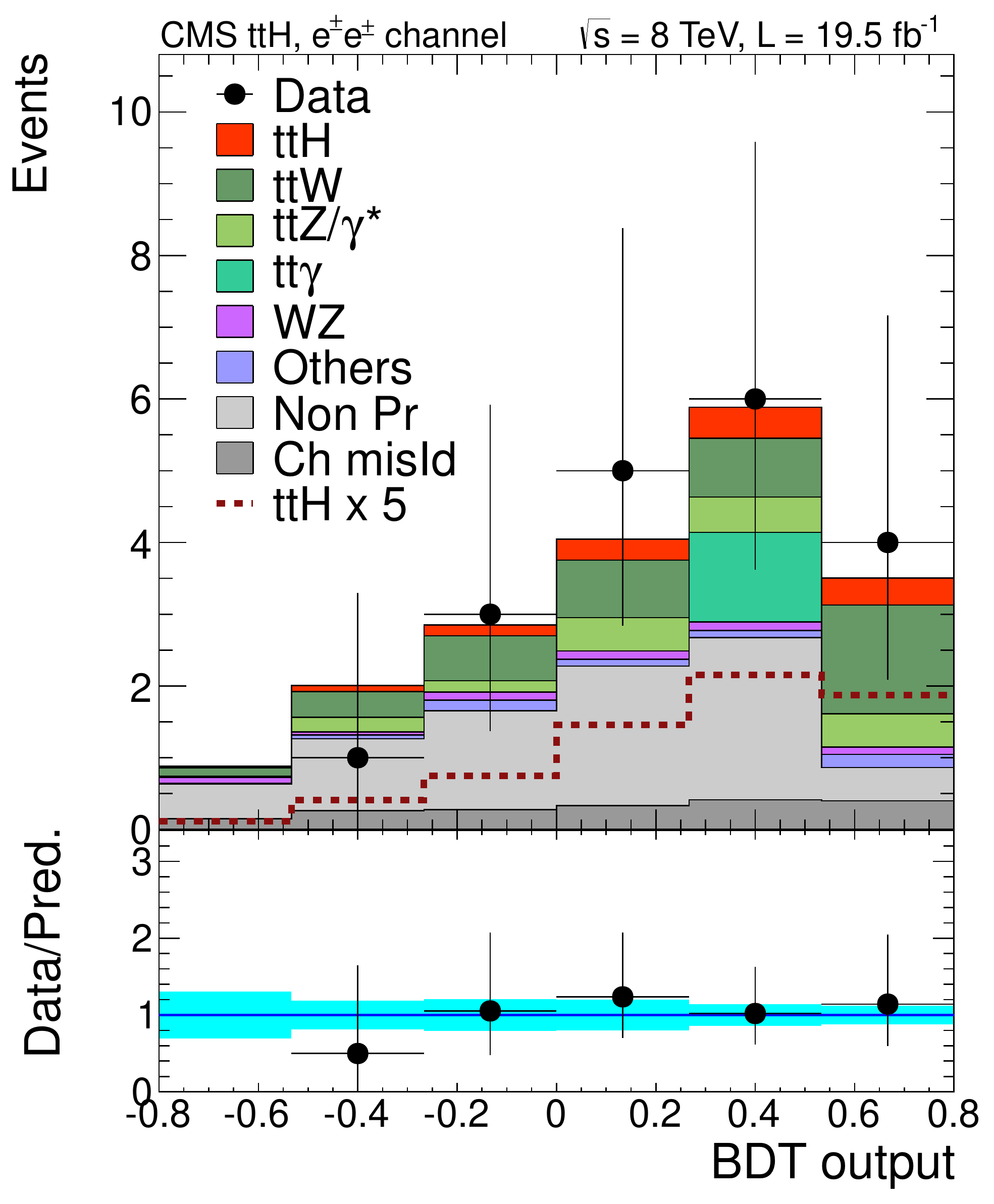}
  \includegraphics[width=0.32\linewidth]{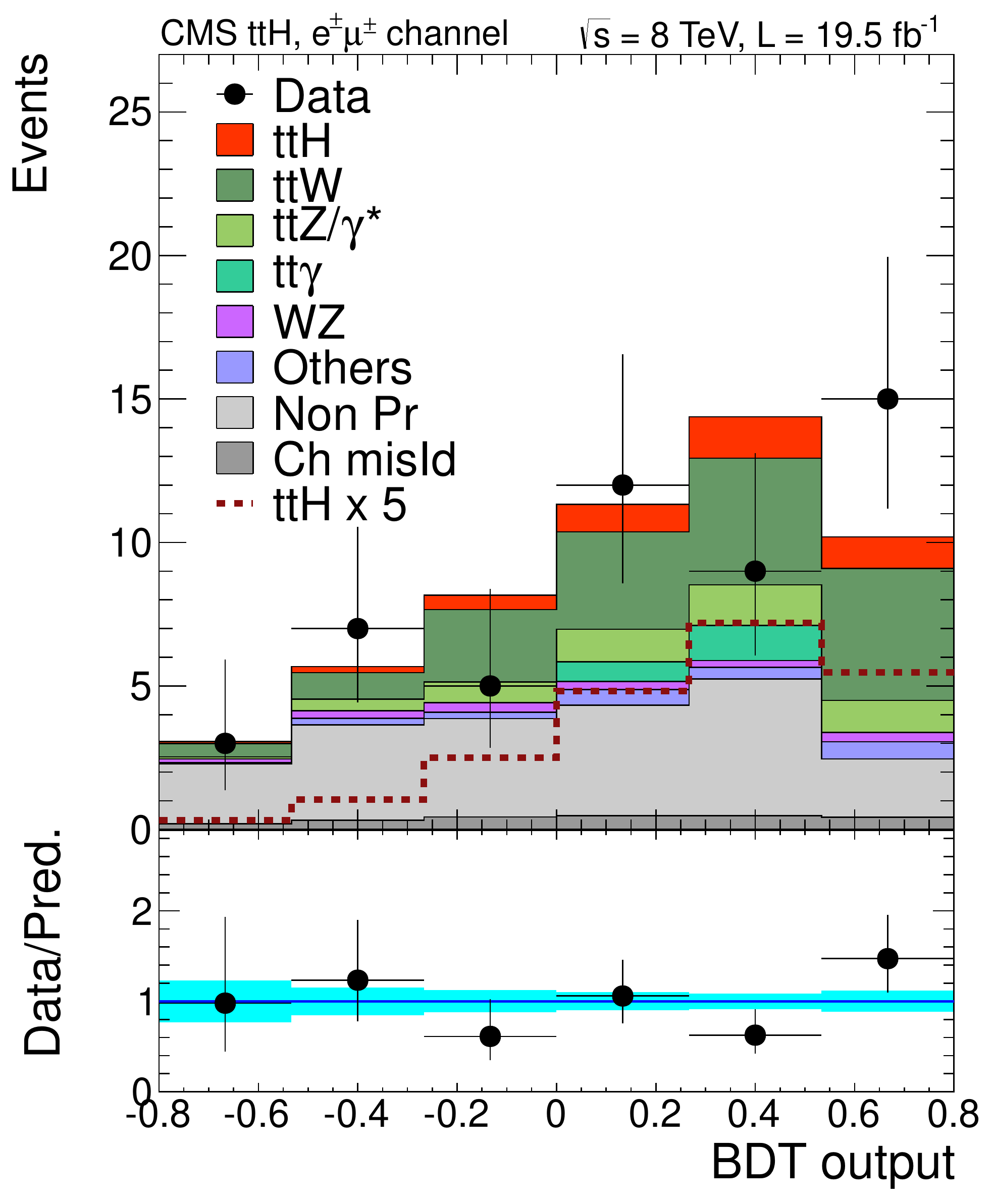}
  \includegraphics[width=0.32\linewidth]{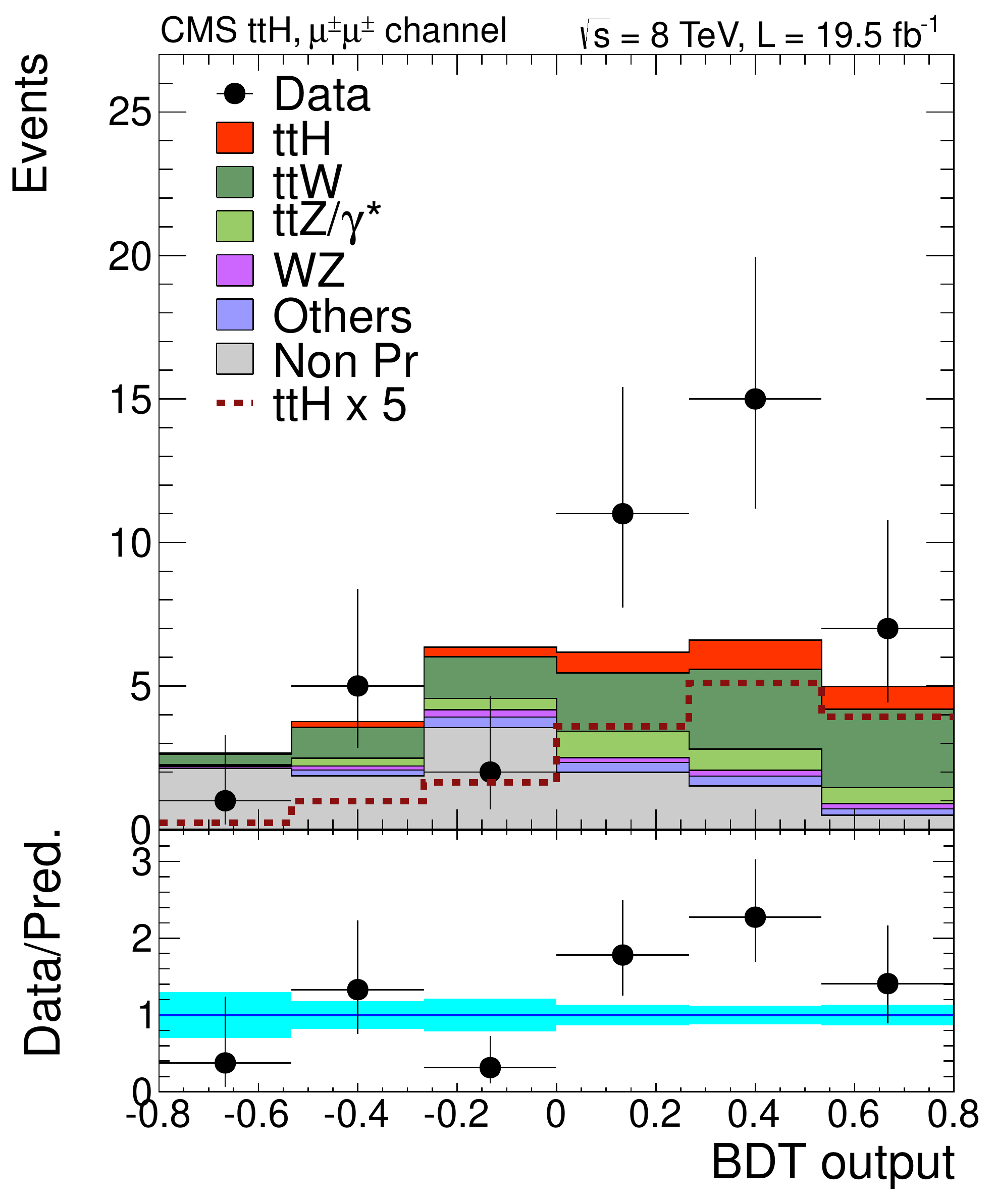}

\caption{Distribution of the final BDT discriminant for the events passing the same-sign di-lepton selection in $ee$ (left), $e\mu$ (center) and $\mu\mu$ (right) channels.}
\label{fig:leptons}
}
\end{figure}

The signal events from $H\rightarrow WW$, $H\rightarrow ZZ$ and $H\rightarrow\tau\tau$ decays contribute to this category with multiple leptons arising as secondary decay products. Selected events are divided to the same-sign dilepton, trilepton, and four-lepton categories and the background contributions are strongly suppressed after the requirement of several leptons along with high jet-multiplicity. Finally, the discriminating variables are combined to a multivariate boosted decision tree (BDT) discriminant to optimally distinguish signal from the background contributions. The final BDT distributions for dilepton analyses are shown in Figure~\ref{fig:leptons}. While $ee$ and $e\mu$ channels show good agreement between observed and expected distributions, a broad signal-like excess of events is visible in the $\mu\mu$ channel. Various checks confirm that this excess does not arise from the mis-modeling of backgrounds.

\section{$t\bar{t}H$ in hadronic final states}
The final state, where the Higgs boson decays to b-quarks has by far the largest cross section. However, the experimental signature suffers from a large multi-jet background of $t\bar{t}$ along with additional jets. The light-flavor component of the $t\bar{t}+ $ jets background can be suppressed by requiring high multiplicity of b-tagged jets in the event. However, $t\bar{t} + b\bar{b}$ contribution remains irreducible, having large theoretical uncertainties \cite{Cascioli:2013era} and a cross-section much larger than that of the signal. Additionally, the presence of b-quarks from top quark decays creates an ambiguity in the Higgs boson mass reconstruction, leading to combinatorial self-background.


\subsection{Search for $t\bar{t}H(\rightarrow b\bar{b})$ and $t\bar{t}H(\rightarrow \tau_{\mathrm{had}}\tau_{\mathrm{had}})$ based on multivariate analysis technique}
\begin{figure}[htbp]
\centering{
  \includegraphics[width=0.4\linewidth]{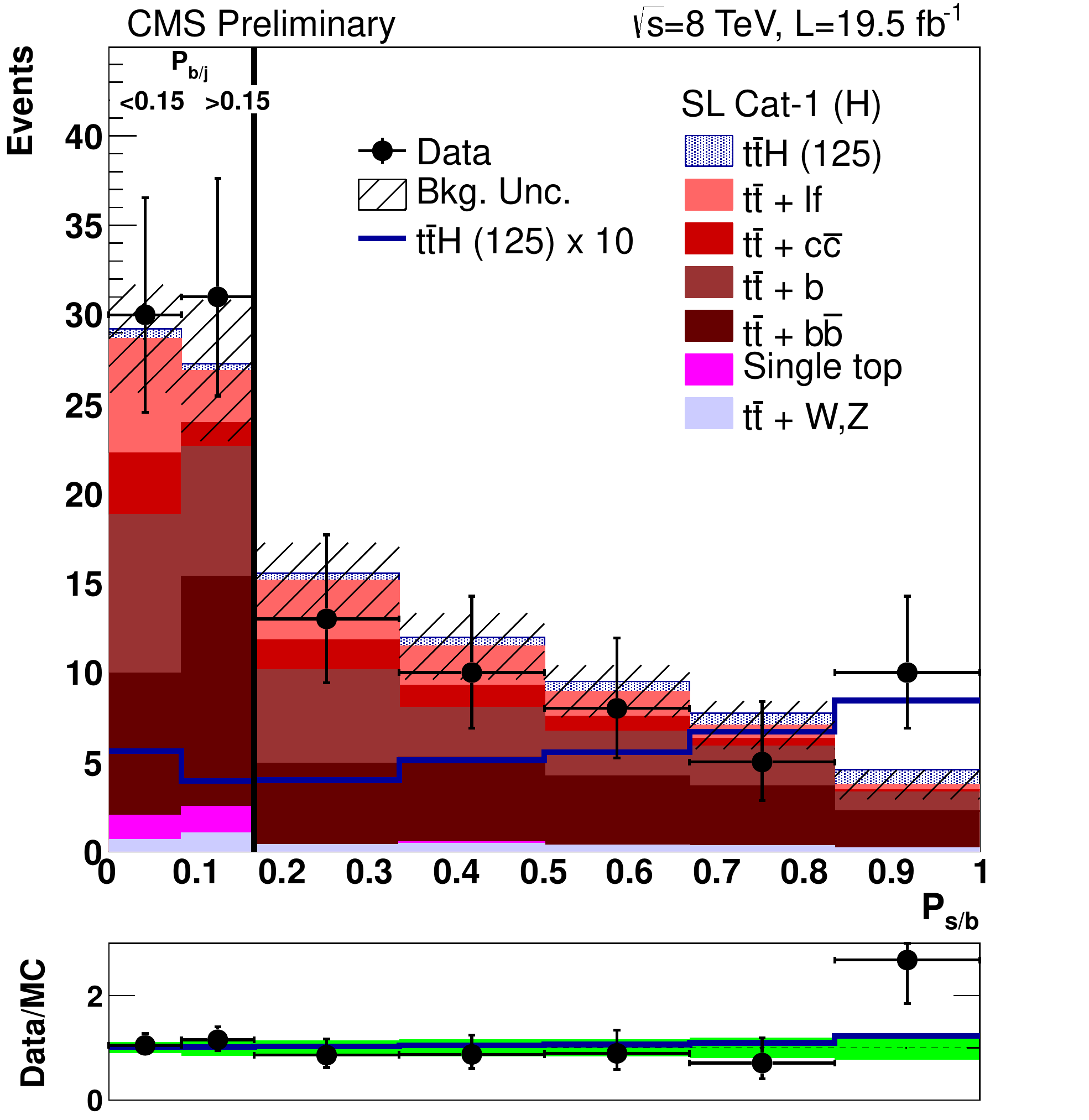}
  \includegraphics[width=0.4\linewidth]{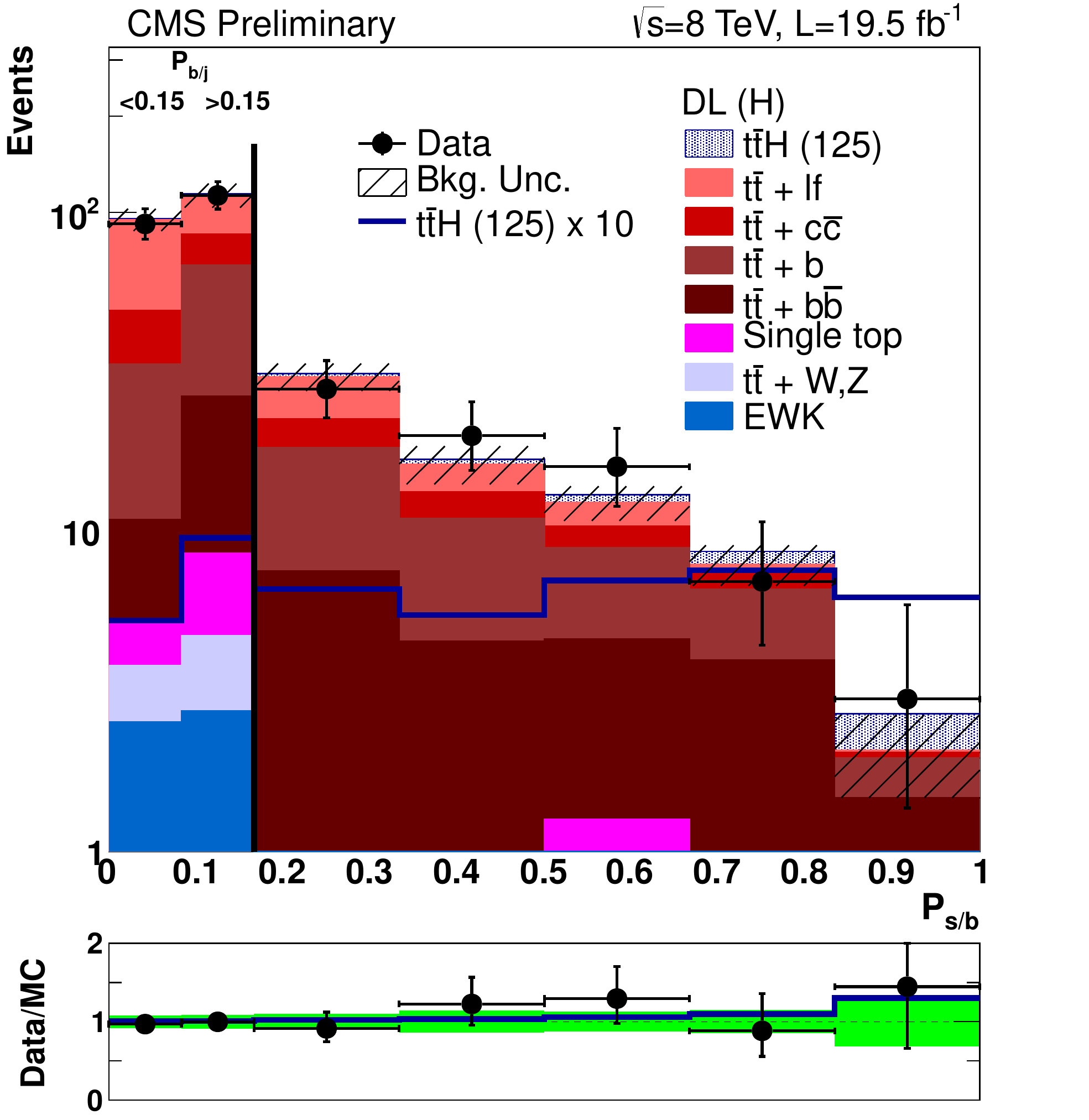}

\caption{Distribution of the MEM discriminant in Cat-1 single lepton events (left) and dilepton event (right). The signal yield is normalized to the standard model prediction. The background yields are obtained from the combined fit to the final discriminant.}
\label{fig:bdthad}
}
\end{figure}

Three different analysis channels are defined, based on the decay modes of $t\bar{t}$ and the Higgs boson: the lepton+jets channel, the dilepton channel, and the $\tau_{\mathrm{had}}$ channel, where only semi-leptonic $t\bar{t}$ decays are considered. Further exclusive categories are defined based on the content of reconstructed jets and b-tagged jets in the event. 
The signal and background fractions for different categories are illustrated in Figure~\ref{fig:bdthad}.
In each category a multivariate BDT discriminant is trained, combining lepton and jet kinematics, event-shape variables and b-tagging information in the event. Finally, a fit to the BDT distribution estimates the number of signal and background contributions. 

\begin{figure}[htbp]
\centering{
  \includegraphics[width=0.55\linewidth]{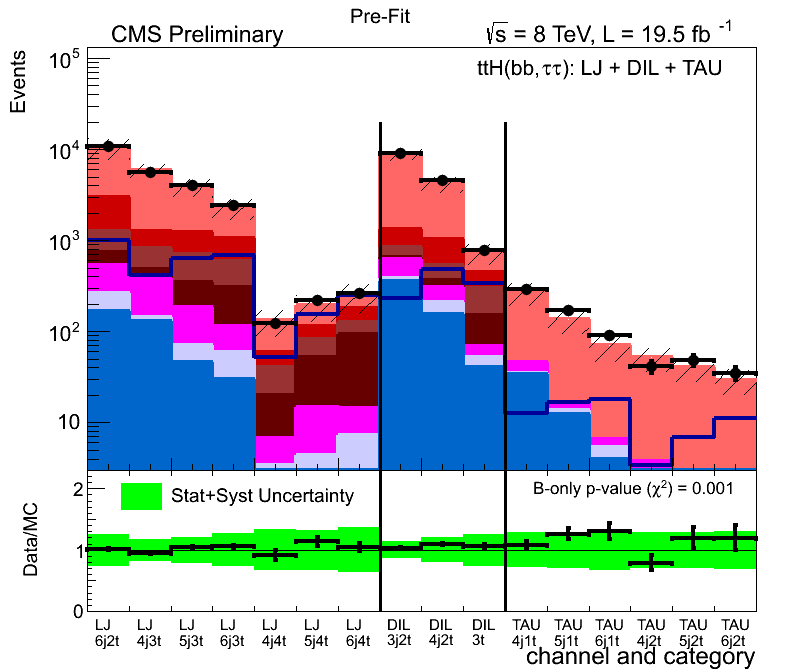}
  \includegraphics[width=0.2\linewidth]{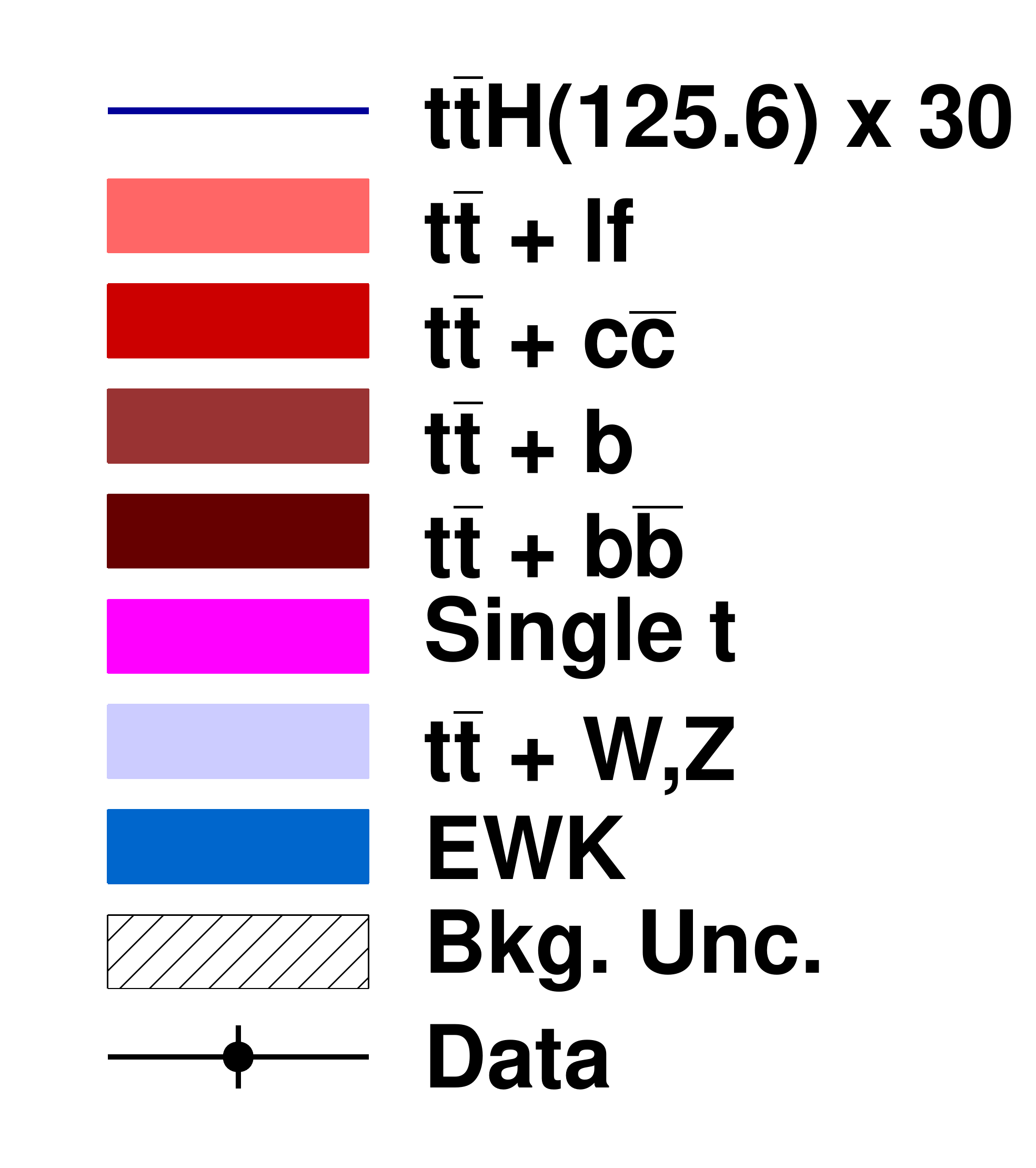}

\caption{Event categorization based on the number of jets and b-tagged jets for $H\rightarrow b\bar{b}$ and $H\rightarrow\tau_{\mathrm{had}}\tau_{\mathrm{had}}$ decay modes.}
\label{fig:mem}
}
\end{figure}

\subsection{Matrix Element Method based search for $t\bar{t}H(\rightarrow b\bar{b})$}
The Matrix Element Method (MEM) \cite{EstradaVigil:2001eq} is a fully analytical multivariate technique that uses the kinematic information in the event without a prior training. The MEM discriminant is obtained by calculating the differential probability density $\omega_i(\mathrm{\mathbf{y}}|\lambda)$ of measuring a set of observables $\mathrm{\mathbf{y}}$ under either a signal ($t\bar{t}H$) or background ($t\bar{t} + b\bar{b}$) hypothesis that depends on unknown model parameters $\lambda$. The ratio of these probability densities provides an optimal separation between signal and background hypotheses. Such approach naturally handles the complicated final state combinatorics, as wrong permutations are assigned low weights. As the method maximally exploits both experimental information and theoretical modeling, it  provides optimal discrimination against the irreducible $t\bar{t} + b\bar{b}$ background. 

After preselecting events by applying a tight requirement on the b-tagged jet content, the events are classified to four exclusive categories based on their parton level event interpretation: three single lepton categories and one dilepton category. Figure~\ref{fig:mem} shows the distributions of MEM discriminant in Cat-1 single lepton channel with events, where all quarks have been reconstructed as jets (left), and in dilepton channel (right). Signal and background yields are obtained from a combined fit in all categories.

\section{Results}
The statistical interpretation of the results follows the same methodology as used for the other CMS Higgs boson analyses \cite{Chatrchyan:2012ufa}. The signal rate is characterized by the signal strength modifier $\mu=\sigma/\sigma_{\mathrm{SM}}$ that is the ratio of the observed $ttH$ production cross section normalized to its standard model expectation.The sensitivity to signal is quantified by the $95\%$ CL upper limit on $\mu$. The results for the individual channels and their combination are shown in Table~\ref{tab:combination}. The combined expected limit for hadronic, leptonic and $\gamma\gamma$ final states (2011+2012 dataset) is $1.7$, nearly reaching the standard model sensitivity. The best-fit value of $\mu$ is measured to be 2.8, showing an excess of events above the background only hypothesis, which largely results from the signal-like excess of events in the same-sign di-muon channel.

\begin{table*}[htbp]

\centering{
\footnotesize{
\begin{tabular}{ l|c|c|c }
\hline
Channel   &  Best-fit $\mu$     & Median exp.  u.l.  &       Observed u.l.  \\  
\hline\hline
\rule[-1.4ex]{0pt}{4ex} $t\bar{t}H(\rightarrow b\bar{b})$           & $0.7^{+1.9}_{-1.9} $ & 3.5  & 4.1 \\
\rule[-1.4ex]{0pt}{4ex} $t\bar{t}H(\rightarrow \tau_{\mathrm{had}}\tau_{\mathrm{had}})$        & $-1.3^{+6.3}_{-5.5}$ & 14.2  & 13.0 \\
\rule[-1.4ex]{0pt}{4ex} $t\bar{t}H(\rightarrow \mathrm{leptons})$        & $3.9^{+1.6}_{-1.4}$ & 2.4  & 6.6 \\
\rule[-1.4ex]{0pt}{4ex} $t\bar{t}H(\rightarrow \mathrm{leptons})$        & $2.7^{+2.6}_{-1.8}$ & 4.7  & 7.4 \\
\hline
\rule[-1.4ex]{0pt}{4ex} Combined & $2.8^{+1.9}_{-0.9}$ & 1.7 & 4.5  \\
\hline
\end{tabular} 
\caption{
The best-fit values of the signal strength modifier obtained from the various individual channels and from their combination.
The observed 95\% CL upper limits on $\mu$ are reported in the third column, and compared to the median expected 
limits for the background only hypothesis.}
\label{tab:combination}
}
}
\end{table*}

The first result of the MEM based analysis in $t\bar{t}H(\rightarrow b\bar{b})$ channel is not yet a part of the combination, but has shown to be extremely promising, already improving the baseline result of $t\bar{t}H(\rightarrow b\bar{b})$ analysis by about 20\% and holding a lot of potential for further improvements.

\end{document}